\documentstyle[12pt]{article}
\title{\bf Tunneling in a Cosmological Model\\
with Violation of Strong Energy Condition}
\vspace{20mm}
\author{M. A. Jafarizadeh$^{a,d}$ \thanks{e-mail: tabriz\_u@vax.ipm.ac.ir} ,
F. Darabi$^{b,c}$ \thanks{e-mail: f-darabi@cc.sbu.ac.ir} ,
A. Rezaei-Aghdam$^{a,c}$ \thanks{e-mail: arezaeia@ark.tabrizu.ac.ir},
A. R. Rastegar$^{a,c,d}$ \thanks{e-mail: rastgar@ark.tabrizu.ac.ir}\\
\\
\\
$^a${\small Department of Theoretical Physics, Tabriz University, Tabriz  51664, Iran.} \\
$^b${\small Department of Physics, Shahid Beheshti University, Tehran 19839, Iran.} \\
$^c${\small Department of Physics, Tarbiyat Moallem University, Tabriz, P.O.Box 51745-406, Iran.} \\
$^d${\small Institute for Studies in Theoretical Physics and Mathematics, Tehran, 19395-1795, Iran.}}
\begin{document}
\maketitle
\vspace{10mm}
\begin{abstract}
The tunneling rate, with exact prefactor, is calculated to first order in $\hbar$
for a closed FRW universe filled with perfect fluid violating the strong energy condition.
The calculations are performed by applying the dilute-instanton approximation on
the corresponding Duru-Kleinert path integral.
It is shown that a closed FRW universe filled with a
perfect fluid with small violation of strong energy condition is more probable
to tunnel than the same universe with large violation of strong energy condition.

\end{abstract}
\newpage
\section{Introduction}

The universe as a whole was the subject of intense investigations in the
early 1960s with the work of DeWitt, Wheeler and Misner \cite{DWM}.
Lately, in early 1980s, the field underwent a rebirth to fully explain the very
early history of universe. The idea of quantum tunneling, responsible for
the birth of the universe, was first introduced by Atkatz and Pagels \cite{AP}
and has been further developed by Vilenkin \cite{V}.
The simplest possible quantum cosmological model is the spontaneous birth from
``nothing'' of an empty FRW universe with a constant vacuum energy density
as its effective cosmological constant. Atkatz and Pagels showed that only a closed
universe can arise via quantum tunneling. The rate of tunneling from ``nothing''
is obtained by WKB approximation for this model \cite{AP}.

The purpose of this paper is to investigate the issue of quantum tunneling for
a closed FRW universe with a perfect fluid source violating the strong energy condition (SEC).
It has been shown \cite{CCS} that neither quantum nor classical wormholes, as instanton solutions
``connecting two asymptotically flat Euclidean domains'',
exist for such a perfect fluid; however the violation of SEC is somehow reasonable in early universe
\cite{CCS} followed by processes such as quantum tunneling.
So we are interested in finding the instantons as ``tunneling solutions'' for this
type of perfect fluid.
To the extent we are concerned with instanton calculations for tunneling rate, this model
may effectively be replaced by an empty closed FRW universe with a cosmological constant.
We shall calculate the tunneling rate by applying the dilute-instanton approximation
to first order in $\hbar$ \cite{CCB}, on the corresponding Duru-Kleinert path integral \cite{JDR}.
Its prefactor is calculated by the heat kernel method \cite{L}, using the
shape invariance symmetry \cite{JF}.

In section ${\bf 2}$, the Duru-Kleinert path integral formula and Duru-Kleinert
equivalence of corresponding actions is briefly reviewed. In section ${\bf 3}$,
we introduce the cosmological model of closed FRW universe with a cosmological
constant as a model without cosmological constant but effectively filled with a perfect fluid
violating SEC. Finally in section ${\bf 4}$, the tunneling rate is fully calculated to first order in
$\hbar$ by applying the dilute-instanton approximation on the corresponding
Duru-Kleinert path integral.

\section{\bf Duru-Kleinert Path Integral}

$\; \; \; \;$ In this section we briefly review the Duru-Kleinert path integral \cite{DK}.
The fundamental object of path integration is the time displacement
amplitude or propagator of a system, $ (X_b \: t_b\: | \: X_a \: t_a) $.
For a system with a time independent Hamiltonian, the object
$ (X_b \: t_b \: | \: X_a \: t_a) $ supplied by a path integral is the causal
propagator
\begin{equation}
(X_b \: t_b \: | \: X_a \: t_a)=\theta(t_a-t_b)<X_b|\exp(-i\hat{H}(t_b-t_a)/\hbar)|X_a>.
\end{equation}
Fourier transforming the causal propagator in the time variable, we
obtain the fixed energy amplitude
\begin{equation}
(X_b \: | \: X_a \: )_E = \int_{t_a}^\infty dt_b e^{iE(t_b-t_a)/\hbar}
(X_b \: t_b\: | \: X_a \: t_a).
\end{equation}
This amplitude contains as much information on the system as the propagator
$(X_b \: t_b\: | \: X_a \: t_a)$, and its path integral form is as follows:
\begin{equation}
(X_b \: | \: X_a)_E = \int_{t_a}^{\infty} dt_b \int {\cal D}x(t) e^{i{\cal A}_E/\hbar}
\label{a}
\end{equation}
with the action
\begin{equation}
{\cal A}_E = \int_{t_a}^{t_b} dt [\frac{M}{2}\dot{x}^2(t)-V(x(t))+E]
\end{equation}
where $ \dot{x} $ denotes the derivatives with respect to $ t $ .
In \cite{DK} it has been shown that fixed energy amplitude (\ref{a}) is equivalent
to the following fixed energy amplitude,
\begin{equation}
(X_b \: | \: X_a)_E = \int_{0}^{\infty} dS [f_r(x_b)f_l(x_a)\int {\cal D}x(s)
e^{i{\cal A}_{E}^{f}/\hbar}]
\end{equation}
with the action
\begin{equation}
{\cal A}_{E}^{f} = \int_{0}^{S} ds \{ \frac{M}{2f(x(s))}x'^2(s)-f(x(s))
[V(x(s))-E] \}
\label{b}
\end{equation}
where $ f_r $ and $ f_l $ are arbitrary regulating functions such that
$f=f_{l} f_{r}$ and $ x'$ denotes the derivatives with respect to time $s$.
The actions $ {\cal A}_E $ and $ {\cal A}_{E}^{f} $,
both of which lead to the same fixed-energy amplitude $ (X_b \: | \: X_a)_E $ are called
Duru-Kleinert equivalent \footnote{Of course a third action
$ {\cal A}_{E,\varepsilon}^{DK} $ is also Duru-Kleinert equivalent of
$ {\cal A}_E $ and $ {\cal A}_E^f $ but we do not consider it here \cite{DK}.}.

In the following section we shall use this equivalence to calculate the quantum
tunneling rate. Thus we rewrite the action $ {\cal A}_{E}^{f} $ in a suitable form such that
it describes a system with zero energy; as only in this sense can we describe
a quantum cosmological model with zero energy.
Imposing $ E = 0 $ in (\ref{b}), with a simple manipulation, gives
\begin{equation}
{\cal A}_{E}^{f} = \int_{0}^{1} ds' S f(X(s')) \{ \frac{M}{2[Sf(X(s'))]^2}
\dot{X}^2(s')-V(X(s')) \}
\end{equation}
where $ \dot{X} $ denotes the derivative with respect to new parameter $ s' $ defined by
\begin{equation}
s' = S^{-1} s
\end{equation}
with $S$ as a  dimensionless scale parameter.\\
After a Wick rotation $ s'=-i\tau $, we get the required Euclidean action and
the path integral
\begin{equation}
I_{0}^{f} = \int_{0}^{1} d\tau S f(X(\tau)) \{ \frac{M}{2[Sf(X(\tau))]^2}
\dot{X}^2(\tau)+V(X(\tau)) \},
\label{z}
\end{equation}
\begin{equation}
(X_b \: | \: X_a) = \int_{0}^{\infty} dS [f_r(X_b)f_l(X_a) \int{\cal D}X(\tau)
e^{{-I_{0}^{f}}/\hbar}]
\end{equation}
where $\tau$ is the Euclidean time. The action (\ref{z}) is Duru-Kleinert equivalent
of
\begin{equation}
I_{0} = \int_{\tau_a}^{\tau_b} d\tau [ \frac{M}{2}\dot{X}^2(\tau)+V(X(\tau)) ]
\end{equation}
where $\tau_a$ and $\tau_b$ correspond to $t_a$ and $t_b$ respectively, and $\dot{X}$ denotes the derivative
with respect to Euclidean time $\tau$.

\section{\bf Model}

We consider an empty closed FRW universe with a nonvanishing cosmological constant
$\Lambda$. The system has only one collective coordinate, namely, the scale factor
$R$. Using the usual Robertson-Walker metric we obtain the scalar curvature
\begin{equation}
{\cal R} = 6 \left[\frac{\ddot{R}}{R}+ \frac{1+{\dot{R}}^2}{R^2} \right].
\label{c}
\end{equation}
Substituting (\ref{c}) into the Einstein-Hilbert action with cosmological constant
leads to the action
\begin{equation}
I=\int_0^1 \! dt \left[-\frac{1}{2}R {\dot{R}}^2+\frac{1}{2}R\left(1-\frac{\Lambda}{3} R^2 \right) \right]
\label{d}
\end{equation}
with the constraint of Einstein equation
\begin{equation}
{\dot{R}}^2+ \left[1-\frac{\Lambda}{3} R^2 \right]=0.
\label{e}
\end{equation}
The system of (\ref{d}) and (\ref{e}) is effectively equivalent  to a model
of closed FRW universe without the cosmological constant but with a stress-energy
tensor of perfect fluid parametrized by constant vacuum energy density $\rho_{vac}$ and pressure
$p=-\rho_{vac}$ \cite{AP}.

Usually a perfect fluid is described by an equation of state such as $p=(\gamma-1)\rho$
where the energy density $\rho(R)$ and
pressure $p(R)$ may be functions of scale factor $R$. We then find, using the
continuity equation, the behaviour of the energy density in a FRW universe \cite{CCS}
\begin{equation}
\rho(R)=\rho(R_0) \left(\frac{R_0}{R}\right)^{3\gamma}
\label{f}
\end{equation}
where $R_0$ is the value of the scale factor at an arbitrary reference time.
It is a well-known fact that SEC is satisfied only if $\gamma>\frac{2}{3}$;
however it is expected to be violated in the early universe \cite{CCS} followed by such
processes as quantum tunneling. So we are interested in investigating the perfect
fluid  for $\gamma\leq\frac{2}{3}$ and look for such processes as quantum
tunneling for the corresponding cosmological model. Thus we rewrite
the relation (\ref{f}) as
\begin{equation}
\rho(R)=\rho(R_0) \left(\frac{R_0}{R}\right)^{2-\frac{2}{\alpha}},
\end{equation}
where $\alpha\geq 0$ indicates the violation of SEC.
As we shall see, to the extent we are concerned only with the action (\ref{d}) and the corresponding
constraint (\ref{e}) dealing with saddle point evaluation of the Euclidean path integral, we may substitute this energy
density into the system (\ref{d}) and (\ref{e}) as the effective cosmological
constant $\Lambda \equiv \rho(R)$  \footnote{We take units such that $\frac{3 \pi}{2 G}=1$.}
\begin{equation}
I=\int_0^1 \! dt \left[-\frac{1}{2}R {\dot{R}}^2+\frac{1}{2}R\left(1-\left(\frac{R}{R_0}\right) ^{\frac{2}{\alpha}} \right) \right],
\label{g}
\end{equation}
\begin{equation}
{\dot{R}}^2+ \left[1-\left(\frac{R}{R_0}\right) ^{\frac{2}{\alpha}} \right]=0
\end{equation}
where $\rho(R_0)=\frac{3}{{R_0}^2}$.
Now, this system describes effectively a closed FRW cosmology in presence of a
perfect fluid violating SEC. The issue of quantum tunneling for this effective
model may be investigated in two ways: WKB approximation, and dilute-instanton
approximation techniques. We shall follow the second approach here, in order to
calculate the tunneling rate $\Gamma$.

\section{\bf Tunneling rate}

It is not suitable to apply the instanton calculation techniques to the Euclidean
form of the action (\ref{g}). The reason is that the kinetic term is
not in its standard quadratic form. It has been recently shown \cite{JDR} that in
such cosmological model one may use the Duru-Kleinert equivalence to work with
the standard form of the action. Using the same procedure, we find the Duru-Kleinert
equivalent action in the cosmological model here as follows
\begin{equation}
I_0=\int_{\tau_a}^{\tau_b}\! dt \left[\frac{1}{2} {\dot{R}(\tau)}^2+\frac{1}{2}R^2\left(1-\left(\frac{R}{R_0}\right) ^{\frac{2}{\alpha}} \right) \right]
\label{h}
\end{equation}
Now, the Euclidean action (\ref{h}) has the right kinetic term to be used in
instanton calculations. The Euclidean type Hamiltonian corresponding to the
action (\ref{h}) is given by
\begin{equation}
H_{E} = \frac{\dot{R}^2}{2} -\frac{1}{2}R^2 \left[1-\left(\frac{R}{R_0}\right) ^{\frac{2}{\alpha}} \right]
\end{equation}
whose vanishing constraint $H_{E}=0$
\footnote{The constraint $H_{E}=0$ corresponds to Euclidean form of the
Einstein equation.} gives a non-trivial instanton solution
\begin{equation}
R(\tau)= \frac{R_0}{(\cosh(\frac{\tau}{\alpha}))^\alpha}
\label{i}
\end{equation}
corresponding to the potential
\begin{equation}
V(R) = \frac{1}{2}R^2 \left[1-\left(\frac{R}{R_0}\right) ^{\frac{2}{\alpha}} \right]\:\:\:\:\:For \:R\geq0.
\label{j}
\end{equation}
Each solution with $\alpha>0$ (violating SEC)
describes a particle rolling down from the top of a potential $ -V(R) $
at $ \tau \rightarrow -\infty $ and $ R = 0 $, bouncing back at $ \tau = 0 $ and
$ R = R_0 $ and finally reaching the top of the potential at $ \tau \rightarrow
+\infty $ and $ R = 0 $.\\
The region of the barrier $ 0 < R < R_0 $ is classically forbidden for the zero energy
particle, but quantum mechanically it can tunnel through it with a tunneling
probability which is calculated using the instanton solution (\ref{i}).\\
The quantized FRW universe is mathematically equivalent to this particle, such
that the particle at $ R = 0 $ and $ R = R_0 $ represents ``nothing'' and ``FRW''
universes respectively. Therefore one can find the probability
$$
|<FRW(R_0) \: | \: nothing>|^2 .
$$
The rate of tunneling $ \Gamma $ is calculated through the dilute instanton
approximation to first order in $\hbar$ as \cite{CCB}
\begin{equation}
\Gamma = [\frac{det'(-\partial_{\tau}^2 + V''(R))}{det(-\partial_{\tau}^2 + \omega^2)}]^{-1/2}
e^{\frac{-I_0(R)}{\hbar}} [\frac{I_0(R)}{2\pi\hbar}]^{1/2}
\label{p}
\end{equation}
where $det'$ is the determinant without the zero eigenvalue,\, $ V''(R) $ is the
second derivative of the potential at the instanton solution (\ref{i}),
$ \omega^2 =V''(R)|_{R=0}$ with $\omega^2=1$ for the potential (\ref{j}),
and $ I_0(R) $ is the corresponding Euclidean action evaluated at the instanton solution (\ref{i}).
The determinant in the numerator is defined as
\begin{equation}
det'[-\partial_{\tau}^2 + V''(R)] \equiv \prod_{n=1}^{\infty}|\lambda_n|
\end{equation}
where $ \lambda_n $ are the non-zero eigenvalues of the operator
$ -\partial_{\tau}^2 + V''(R) $.\\
The explicit form of this operator is obtained as
\begin{equation}
O \equiv \alpha^{-2}[-\frac{d^2}{dx^2} - \frac{(\alpha+1)(\alpha+2)}{\cosh^2 x}+\alpha^2]
\label{k}
\end{equation}
where we have used (\ref{i}) and (\ref{j}) with a change of variable $ x = \frac{\tau}{\alpha} $.
Now, in order to find exactly the
eigenvalues and eigenfunctions of the operator (\ref{k}) we assume $\alpha$ to be
positive integer. By relabeling $l=\alpha+1$, the eigenvalue equation
of the operator (\ref{k}) can be written as
\begin{equation}
\Delta_l\psi_l(x)=(E_l-2l+1)\psi_l(x)
\label{y}
\end{equation}
with
\begin{equation}
\Delta_l:=-\frac{d^2}{dx^2}-\frac{l(l+1)}{\cosh^2 x}+l^2
\end{equation}
where the factor $\alpha^{-2}$ is ignored for the moment. The equation (\ref{y}) is a
time independent Schrodinger equation.
Now, by ignoring the constant shift of energy $2l-1$ and by
introducing the following first order differential operators
\begin{equation}
\left\{\begin{array}{ccc}B_l(x): \:=\:\frac{d}{dx}+l\tanh{x} \\
B^\dagger_l(x): \:=\:-\frac{d}{dx}+l\tanh{x},\end{array}
\right.
\end{equation}
the operator $\Delta_l$ can be factorized and
using the shape invariance symmetry we have \cite{JF}
\begin{equation}
\psi_l(x)= \frac{1}{\sqrt{E_l}}B^\dagger_l(x)\psi_{l-1}(x)
\end{equation}
\begin{equation}
\psi_{l-1}(x)= \frac{1}{\sqrt{E_l}}B_l(x)\psi_l(x)
\end{equation}
Therefore, for a given $l$, its first (bounded) excited state can be obtained from
the ground state of $l-1$.
Consequently, the excited state $m$ of a given $l$, that is $\psi_{l,m}$,
can be written as
\begin{equation}
\psi_{l,m}(x)=\sqrt{\frac{2(2m-1)!}{\Pi^m_{j=1} j(2l-j)}}\frac{1}{2^m(m-1)!}
B^\dagger_l(x)B^\dagger_{l-1}(x) \cdots B^\dagger_{m+1}(x) \frac{1}{\cosh^m x},
\end{equation}
with eigenvalues $E_{l,m}=l^2-m^2$.
Also its continuous spectrum consists of
\begin{equation}
\psi_{l,k}=\frac{B_{l}^\dagger(x)}{\sqrt{k^2+l^2}}\frac{B_{l-1}^\dagger(x)}{\sqrt{k^2+(l-1)^2}} \cdots
\frac{B_{1}^\dagger(x)}{\sqrt{k^2+1^2}}\frac{e^{ikx}}{\sqrt{2\pi}},
\label{t}
\end{equation}
with eigenvalues $E_{l,k}=l^2+k^2$ where $\int^{+\infty}_{-\infty}\psi^{*}_{l,k}(x)\psi_{l,k^{\prime}}(x)dx
=\delta(k-k^{\prime})$.
Now, we can calculate the ratio of the determinants as follows.
First we explain very briefly how one can calculate the determinant of an
operator by the heat kernel method \cite{L}. We introduce the generalized
Riemann zeta function of the operator $A$ by
\begin{equation}
\zeta_A(s) = \sum_{m} \frac{1}{|\lambda_m|^s},
\label{m}
\end{equation}
where $ \lambda_m $ are eigenvalues of the operator $A$,\, and the determinant
of the operator $A$ is given by
\begin{equation}
det \, A = e^{-\zeta'_{A}(0)}.
\label{n}
\end{equation}
It is obvious from the equations (\ref{m}) and (\ref{n}) that
for an arbitrary constant $c$
\begin{equation}
det(c A)=c^{\zeta_A(0)}det A .
\label{r}
\end{equation}
On the other hand $ \zeta_A(s) $ is the Mellin transformation of the heat kernel
$ G(x,\, y,\, \tau)$
\footnote{Here $\tau$ is a typical time parameter.}
which satisfies the following heat diffusion equation
\begin{equation}
A \, G(x,\,y,\, \tau) = -\frac{\partial \, G(x,\,y,\, \tau)}{\partial \tau},
\label{o}
\end{equation}
with an initial condition $ G(x,\,y,\,0) = \delta(x - y) $.\, Note that
$ G(x,\,y,\, \tau) $ can be written in terms of its spectrum
\begin{equation}
G(x,\,y,\, \tau) =  \sum_{m} e^{-\lambda_{m}\tau} \psi_{m}^{*}(x) \psi_{m}(y).
\end{equation}
An integral is written for the sum if the spectrum is continuous.
From relations (\ref{n}) and (\ref{o}) it is clear that
\begin{equation}
\zeta_{A}(s) = \frac{1}{\Gamma(s)} \int_{0}^{\infty} d\tau \, \tau^{s-1}
\int_{-\infty}^{+\infty} dx \, G(x,\,x,\, \tau).
\label{w}
\end{equation}
Now, in order to calculate the ratio of the determinants in (\ref{p}), called a
prefactor, we need to find the difference of the functions
$ G(x, y, \tau) $ for two operators $\Delta_l , \Delta_l(0)$, where
\begin{equation}
\Delta_l(0):=-\frac{d^2}{dx^2}+l^2 .
\end{equation}
Considering the fact that $\Delta_l+1-2l$ (or $\Delta_l(0)+1-2l$) has the
same eigen-spaces as $\Delta_l$ (or $\Delta_l(0)$) and the eigen-spectrum
is shifted by $1-2l$, we have
\begin{equation}
G_{\Delta_l(0)+1-2l}(x,y,\tau)=\frac{e^{-(l-1)^2 \tau}}{2\sqrt{\pi\tau}}e^{-\frac{(x-y)^2}{4\tau}}
\end{equation}
\begin{eqnarray}
\nonumber
G_{\Delta_l+1-2l}(x,y,\tau)=\sum^{l-1}_{m=0,m\neq 1}\psi^{*}_{l,m}(x)\psi_{l,m}(y)e^{-|(m-1)(2l-(m+1))|\tau}+\\
 \int^{+\infty}_{-\infty}dke^{-((l-1)^2+k^2)\tau}\psi^{*}_{l,k}(x)\psi_{l,k}(y).
\end{eqnarray}
In order to  calculate the function $\zeta_{\Delta_l+1-2l}$, according
to the relation (\ref{w}) we have to take the trace of heat kernel $G_{\Delta_l+1-2l}(x,y,\tau)$
where we need to integrate over $|\psi_{l,k}|^2$. Using the relation
$\frac{B_l}{\sqrt{E_{l,k}}}\psi_{l,k}(x)=\psi_{l-1,k}(x) $ we have
\begin{eqnarray}
\nonumber
\int^{+\infty}_{-\infty}dx\psi^{*}_{l,k}(x)\psi_{l,k}(x)=-\lim_{x\rightarrow\infty}\frac{1}{\sqrt(E_{l,k})}\psi^{*}_{l,k}(x)\psi_{l-1,k}(x) +\\
\lim_{x\rightarrow -\infty}\frac{1}{\sqrt(E_{l,k})}\psi^{*}_{l,k}(x)\psi_{l-1,k}(x) +\int^{+\infty}_{-\infty}dx\psi^{*}_{l-1,k}(x)\psi_{l-1,k}(x).
\label{u}
\end{eqnarray}
The first and the second terms appearing on the right hand side of the recursion
relation (\ref{u}) are proportional to the asymptotic value of the wave functions at
$ \infty $ and $-\infty $, respectively, where the latter is calculated as
\begin{eqnarray}
\nonumber
\lim_{x\rightarrow \pm\infty} \psi_{m,k}(x)=\frac{-ik\pm m }{\sqrt{k^2+m^2}}\frac{-ik\pm (m-1) }{\sqrt{k^2+(m-1)^2}}\cdots \frac{-ik\pm 1 }{\sqrt{k^2+1}}\frac{\exp(ikx)}{\sqrt{2\pi}}=\\
\frac{1}{\sqrt{2\pi}}\prod^m_{j=1}(\frac{-ik\pm j }{\sqrt{k^2+j^2}})\exp(ikx).
\end{eqnarray}
Substituting these asymptotic behaviours in the recursion
relations between the norms of the wave functions $\psi_{m,k}$
associated with the continuous spectrum (\ref{t}), then using the obtained
recursion relations together with the orthonormality of discrete spectrum
we get the following result for the difference of traces of heat kernels:
\begin{eqnarray}
\nonumber
\int^{+\infty}_{-\infty}dx(G_{\Delta_l+1-2l}(x,x,\tau)-G_{\Delta_l(0)+1-2l}(x,x,\tau))=\;\;\;\;\;\;\;\;\;\;\;\;\;\;\;\;\;\;\;\;\;\;\;\;\;\;\;\;  \\
\nonumber
\sum_{m=0,m \neq 1}^{l-1}\exp(-|(m-1)(2l-(m+1))|\tau)-\frac{1}{\pi}\sum_{m=1}^{l}m(\int^{+\infty}_{-\infty}dk\frac{\exp(-((l-1)^2+k^2)\tau}{(k^2+(l-1)^2)}\\
+((l-1)^2-m^2)\int^{+\infty}_{-\infty}dk\frac{\exp(-((l-1)^2+k^2)}{(k^2+m^2)(k^2+(l-1)^2)})\;\;\;\;\;\;\;\;\;\;\;\;\;\;\;\;\;\;\;\;\;\;\;\;\;\;\;\;\;\;\; .
\label{s}
\end{eqnarray}
Hence, using the Mellin transformation (\ref{w}) and the well-known Feynman integral
\begin{eqnarray}
\nonumber
\hspace{-100mm} \frac{1}{D_1^{a_1}D_2^{a_2} \cdots D_n^{a_n}}=\;\;\;\;\;\;\;\;\;\;\;\;\;\;\;\;\;\;\;\;\;\;\;\;\;\;\;\;\;\;\;\;\;\;\;\;\;\;\;\;\;\;\;\;\;\;\\
\frac{\Gamma(a_1+a_2+ \cdots +a_n)}
{\alpha(a_1)\Gamma(a_2) \cdots \Gamma(a_n)}\int dt_1dt_2 \cdots dt_n\frac
{\delta(1-t_1-t_2 \cdots -t_n)t_1^{a_1-1}t_2^{a_2-1} \cdots t_n^{a_n-1}}
{(t_1D_1+t_2D_2+ \cdots +t_nD_n)^{a_1+a_2+ \cdots +a_n}}.
\end{eqnarray}
We finally get
\begin{eqnarray}
\nonumber
\zeta_{\Delta_l+1-2l}(s)-\zeta_{\Delta_l(0)+1-2l}(s)=\;\;\;\;\;\;\;\;\;\;\;\;\;\;\;\;\;\;\;\;\;\;\;\;\;\;\;\;\;\;\;\;\;\;\;\;\;\;\;\;\;\;\;\;\;\;\; \\
\nonumber
\sum_{m=0,m \neq 1}^{l-1}(|(m-1)(2l-(m+1))|)^{-s}-\frac{1}{2\pi}l(l+1)(l-1)^{-(2s+1)}\beta(s+\frac{1}{2},\frac{1}{2})\;\;\;\;\;\;\;\;\;\;\;\;\;\;\;\;\;\;\;\;\; \\
\nonumber
-\frac{1}{\sqrt{\pi}}\frac{\Gamma(s+\frac{3}{2})}{\Gamma(s+2)}\sum_{m=1}^{l-1}m(l-1)^{-(2s+3)}((l-1)^2-m^2)_2F_1(s+\frac{3}{2},1,s+2,1-\frac{m^2}{(l-1)^2})\\
\nonumber
\hspace{-80mm}-\frac{1}{\sqrt{\pi}}\frac{\Gamma(s+\frac{3}{2})}{\Gamma(s+2)}l^{-2(s+1)(1-2l)}_2F_1(s+\frac{3}{2},s+1,s+2,1-\frac{(l-1)^2}{l^2})\;\;\;\;\;\; \\
\label{v}
\end{eqnarray}
where $\beta$ is the $beta$ function.\\
For $s=0$, we obtain
\begin{equation}
\zeta_{\Delta_l+1-2l}(s)-\zeta_{\Delta_l(0)+1-2l}(s)|_{s=0}=-1.
\label{q}
\end{equation}
This means that the operators $ \Delta_l+1-2l $  and $ \Delta_l(0)+1-2l $
have the same number of eigen-spaces (even though  for both of them this
number is infinite) since from the definition of Riemann's zeta function, it is
obvious that its value at $s=0$ can be interpreted as  the number of eigen
spaces of the corresponding operator. The appearance of $-1$ on the right hand
side of the relation is due to the ignorance of the eigen-functions associated
with the zero eigen value of the operator $ \Delta_l+1-2l $. Therefore, its
number of eigen-states is the same as that of the operator $ \Delta_l(0)+1-2l $.
In order to calculate the ratio of the determinant of the operators
$ \Delta_l+1-2l $  and $ \Delta_l(0)+1-2l $ we need to know the derivative
of their associated zeta functions at $s=0$. Hence differentiating both sides of the
relation (\ref{q}) with respect to $s$ and evaluating such integrals as
\begin{equation}
\int^1_0dt(((l-1)^2-m^2)t+m^2)^{-\frac{3}{2}}\log t=\frac{4}{m((l-1)^2-m^2)}(\log 2+\log m-2\log (l-1+m)),
\end{equation}
\begin{eqnarray}
\nonumber
\hspace{-10mm}\int^1_0dt(((l-1)^2-m^2)t+m^2)^{-\frac{3}{2}}\log (((l-1)^2-m^2)t+m^2)\log t=\\
\hspace{-20mm}\frac{4}{m((l-1)^2-m^2)}(\log m\frac{m}{l-1}\log(l-1)+\frac{(l-1-m)}{l-1}) m-2\log (l-1+m)),
\end{eqnarray}
we get
\begin{equation}
\zeta^{\prime}_{\Delta_l+1-2l}(s)-\zeta^{\prime}_{\Delta_l(0)+1-2l}(s)|_{s=0}=\log (\frac{2(2l-1)!}{((l-2)!)^2}.
\label{2}
\end{equation}
Therefore, according to the relations (\ref{p}), (\ref{n}), (\ref{r}), (\ref{q}) and (\ref{2}) the prefactor
associated with the potential (\ref{j}), is calculated as
\begin{equation}
\frac{det(\frac{1}{(l-1)^2}(-\frac{d^2}{dx^2}-\frac{l(l+1)}{\cosh^2x}+(l-1)^2))}
{det(\frac{1}{(l-1)^2}(-\frac{d^2}{dx^2}+(l-1)^2))}=\frac{((l-1)!)^2}{2(2l-1)!}.
\end{equation}
Finally, the decay rate of metastable state of this potential is calculated as
\begin{equation}
\Gamma=\frac{1}{\sqrt{\pi\hbar}}R_0 2^{l-1}\exp(-\frac{(l-1)R_0^2 \beta(\frac{3}{2},l-1)}{\hbar})+O(\hbar)
\label{1}
\end{equation}
where we have used the value of $I_0$ at the instanton solution (\ref{i}):
\begin{equation}
I_0=\int_{-\infty}^{\infty}R_0^2 \frac{(\sinh(\frac{\tau}{\alpha}))^2}{(\cosh(\frac{\tau}{\alpha}))^{2(1+\alpha)}}d\tau=\alpha R_0^2 \beta(\frac{3}{2},\alpha).
\end{equation}
The result (\ref{1}) agrees with one obtained by the WKB approximation, and generalizes the decay rate
obtained for $l=2 $ (or $\alpha=1$) in \cite{JDR}.\\
\\
\\
{\Large \it {\bf Concluding Remarks}}
\\
\\
In this paper we have calculated the tunneling rate, with exact prefactor, to
first order in $\hbar$ from ``nothing'' to a closed FRW universe filled with
perfect fluid violating the strong energy condition. We were motivated by the
belief that the strong energy condition is expected to be violated in the early
universe, especially in such processes as quantum tunneling. The result obtained
here somehow confirms this fact in that the tunneling rate (\ref{1}) increases
for higher values of the positive integer $\alpha$ (or $l$) such that for
$\alpha \rightarrow\infty$ the very high tunneling rate corresponds to the
point $\gamma=2/3$ below which the strong energy condition is violated \cite{CCS}.
On the other hand for $\alpha=1$ the tunneling rate has its minimum value corresponding
to $\gamma=0$ (cosmological constant \cite{CCS}) as more violation of strong energy condition.
In this respect, the main result is that a closed FRW universe filled with a
perfect fluid with small violation of strong energy condition is more probable
to tunnel than the same universe with large violation of strong energy condition.

\end{document}